\documentclass[11pt]{article}
\usepackage{graphicx,amsfonts,amsthm,amssymb}
\theoremstyle{plain}
\newtheorem{thm}{{\bf Theorem}}[section]
\newtheorem{lemma}{{\bf Lemma}}[section]
\newtheorem{prop}{{\bf Proposition}}[section]
\newtheorem{cor}{{\bf Corollary}}[section]
\newtheorem{claim}{{\bf Claim}}[section]

\theoremstyle{remark}
\newtheorem{remark}{{\it Remark}}[section]
\newtheorem{ex}{{\it Example}}[section]

\def\be{\begin{eqnarray}}
\def\ee{\end{eqnarray}}
\def\ben{\begin{eqnarray*}}
\def\een{\end{eqnarray*}}
\def\ba{\begin{array}}
\def\ea{\end{array}}
\def\bp{\noindent{\it Proof. }}
\def\ep{\noindent{\hfill \fbox{}}}

\def\pic{{\rm Pic}}

\def\rank{{\rm rank}}
\def\noi{\noindent}
\def\nn{\nonumber}
\def\vp{\varphi}
\def\al{\alpha}
\def\ve{\varepsilon}

\def\pone{{\mathbb P}^1}
\def\ptwo{{\mathbb P}^2}
\def\pth{{\mathbb P}^3}
\def\ponet{{\mathbb P}^1\times{\mathbb P}^1}
\def\mc{{\mathbb C}}
\def\mz{{\mathbb Z}}
\def\mq{{\mathbb Q}}

\def\nn{\nonumber}

\newcommand{\ol}[1]{\overline{#1}}

\newcommand{\mapright}[1]{%
   \smash{\mathop{%
   \hbox to 1cm{\rightarrowfill}}\limits^{#1}}}
\newcommand{\mapleft}[1]{%
   \smash{\mathop{%
   \hbox to 1cm{\leftarrowfill}}\limits^{#1}}}
\newcommand{\maplleft}[2]{%
   \smash{\mathop{%
   \hbox to 1cm{\leftarrowfill}}\limits_{#1}^{#2}}}

\begin{document}

\title{Discrete dynamical systems associated with 
the configuration space of 8 points in ${\mathbb P}^3({\mathbb C})$}
\author{Tomoyuki Takenawa}
\date{}
\maketitle

\begin{center}
Graduate School of Mathematical Sciences, University of Tokyo\\ 
Komaba 3-8-1, Meguro-ku, Tokyo 153-8914, Japan\\
\end{center}

\begin{abstract} 
A 3 dimensional analogue of Sakai's theory concerning the relation 
between rational surfaces and discrete Painlev\'e equations 
is studied. For a family of rational varieties obtained by blow-ups 
at 8 points in general position in ${\mathbb P}^3$, 
we define its  symmetry group 
using the inner product that is associated with the intersection numbers and 
show that the group is isomorphic to the Weyl group of type $E_7^{(1)}$. 
By normalizing the configuration space by means of elliptic curves, 
the action of the Weyl group and the dynamical system associated with 
a translation are explicitly described. 
As a result, it is found that the action of the Weyl group on 
${\mathbb P}^3$ preserves a one parameter family of quadratic surfaces 
and that it can therefore be reduced to the action on 
${\mathbb P}^1\times {\mathbb P}^1$. 
\end{abstract}

\section{Introduction} 
Relations between Painlev\'e equations and rational surfaces 
were first studied by Okamoto\cite{okamoto}. He showed that 
for each Painlev\'e equation, by elimination of the singularity
of the equation, the solutions can be regularly 
extended into a family of rational surfaces. 
Such a family of rational surfaces is called {\it the space of initial 
conditions} 
for the Painlev\'e equation. 
Conversely, it is clarified by Saito and Takano et al. \cite{st,stt} that 
for a given space of initial conditions 
the Hamilton system of a Painlev\'e equation can be determined. 

Since the singularity confinement method was introduced by Grammaticos et al. 
\cite{grp}, the discrete Painlev\'e equations have been studied 
extensively (\cite{rgh} for example). 
Emphasizing the fact that each discrete Painlev\'e equation 
preserves a family of rational surfaces, Sakai 
constructed the discrete Painlev\'e equations 
from families of rational surfaces (called generalized Halphen surfaces) 
and subsequently classified them. 
Such a family of rational surfaces is called {\it the space of initial 
conditions} for that discrete Painlev\'e equation. 
A generalized Halphen surface can be seen to be isomorphic to a surface 
obtained by 9 blow ups from $\ptwo$. 
Sakai's classification also shows that the spaces of initial conditions 
for the discrete Painlev\'e equations include those for continuous ones. 
Furthermore, the largest symmetry 
arises when the 9 points are in general position; all other symmetries
are in the case where the points are in some special position. 
Each Painlev\'e equation is obtained as a translation associated with
the corresponding affine Weyl group 
(extended by the automorphisms of the associated Dynkin 
diagram). Moreover, if the space of initial conditions is that of a 
continuous one, its Weyl group coincides with the group 
of that equation's B\"acklund transformations. 

The aforementioned results all concern non-autonomous dynamical systems. 
There also exist some studies that deal with autonomous ones. 
For example, in the continuous case Adler and van Moerbeke 
have studied Painlev\'e manifolds \cite{am} and in the discrete case 
various authors have studied the relations between dynamical systems and 
the automorphism groups of manifolds (\cite{df,gizatullin} for example). 

Sakai's procedure for describing discrete Painlev\'e equations is 
closely related to 
the studies on the Cremona isometry carried out by Coble et al. 
\cite{coble,do} 
and these two approaches coincide in the case of 9 points in general position. 
Whereas in the case of points in general position the Weyl group is generated 
by the standard Cremona transformation and exchanges of the points, 
in the degenerate case its generators can be constructed by changing
the blow down structures. Concerning this point one has to cite the 
pioneering research by Looijenga \cite{looijenga}. 

Dolgachev and Ortland also studied the case of 
3 (or more) -dimensional rational 
varieties \cite{do}: 
here (for example) the affine Weyl group of type $E_7^{(1)}$ 
appears in the case where 8 generic points in $\pth$ are blown-up. 
If the number of points is less than 8 the Weyl group is finite and 
it is indefinite if the number of points is larger than 8. 
However, in the 3-dimensional case the action of each element 
of the Weyl group cannot be 
lifted to an isomorphism 
between rational varieties, obtained by blow-ups at some points. 
Dolgachev and Ortland call such a map a pseudo-isomorphism. Note that although 
they described the roots of 
the Weyl group by means of a Neron-Severi bi-lattice, 
the geometric origin of the dual bases were not clear. 

In this paper we study the symmetry, the normalization 
of the configuration space and the associated discrete dynamical systems 
for the family of rational varieties obtained by blow-up at 8 ordered points 
in general position in $\pth$. 

In section 2, we reconstruct the argument of Dolgachev and Ortland. 
We consider birational automorphisms of the family of varieties
such that (i) each of them acts as an automorphism 
on the configuration space (ii) for rational varieties on the 
configuration space it preserves the ``inner product'' 
of the Picard group $\pic(X)$. Here, the inner product is
defined by using the intersection numbers and
the canonical divisor $K_X$ as
$(D,D'):=D\cdot D'\cdot (-\frac{1}{2}K_X)$ for $D,D' \in \pic(X)$.
It is shown that the resulting symmetry group is the Weyl group 
of type $E_7^{(1)}$. 
This group coincides with that of 
Dolgachev and Ortland.

In section 3, the normalization of the configuration space is discussed. 
Although there is a straightforward normalization, using this it is difficult 
to describe the action of the Weyl group and to see the properties 
of the resulting dynamical systems. 
In this paper we therefore use a normalization in terms of elliptic curves. 
Quadratic surfaces passing through the 8 points we consider form 
at least one parameter family. Here the 8 points are on the intersection 
curve of the pencil of surfaces.
Normalizing the pencil, one obtains a 
normalization of the configuration space. 

In section 4, we describe the action of the Weyl group obtained in section 2 
in normalized coordinates. In order to calculate the concrete action 
we apply a 3-dimensional analogue of the period map which is
introduced by Looijenga-Sakai for surfaces. 

In section 5, we construct a birational dynamical system 
in $\pth$ by using the action obtained in section 4. 
Such systems are obtained corresponding to translations 
associated to the Weyl group. We describe one of them explicitly.

Is section 6, it is shown that the action of the Weyl group 
preserves each element of the pencil of quadratic surfaces used for the
normalization and that it can therefore be reduced to an action on $\ponet$. 
The reduced action of the Weyl group of type $E_7^{(1)}$ on $\ponet$ 
coincides with the action of a sub-group of the Weyl group 
of type $E_8^{(1)}$, 
which is the symmetry of the family of (the most) general Halphen surfaces. 

Section 7 is devoted to conclusions and discussions.

\section{Symmetry}

Let $X(4,8)$ denote the configuration space of ordered 8 points in $\pth(\mc)$
such that every 4 points are not on the same plane:
\be && \hspace{-0.5cm}
PGL(4,\mc) \left\backslash \left\{
\left. \left(\ba{cccc}
x_1&x_1&\cdots&x_8\\
y_1&y_2&\cdots&y_8\\
z_1&z_3&\cdots&z_8\\
w_1&w_4&\cdots&w_8\ea\right) \in (\mc)^8 \right| 
\ba{l} \mbox{every $4\times 4$}\\ \mbox{minor determinant}\\
\mbox{is nonzero}\ea  \right\}
 \right/ (\mc^{\times})^8 \label{x8},
\ee
where two configurations are identified if one can be transformed
to the other by a projective transformation. 
We also denote the 3-dimensional rational variety obtained by successive 
blowing-up at distinct 8 points $P_i(x_i:y_i:z_i:w_i)$ 
by $X_{P_1,\cdots,P_8}$ (or simply by $X$) and the family
of all $X_{P_1,\cdots,P_8}$'s, where $\{P_1,\cdots,P_8\}\in X(4,8)$,
by $\{X_{P_1,\cdots,P_8}\}$. $X(4,8)$ is called the parameter space.

Let $\pic(X)$ be the Picard group of the variety $X$ (the additive
group of isomorphism classes of invertible sheaves $\simeq$
the additive group of linear equivalence classes of divisors).
We have
\ben
\pic(X)&=& \mz E\oplus \mz E_1\oplus \mz E_2\oplus \mz E_3\oplus 
\mz E_4\oplus \mz E_5\oplus \mz E_6\oplus \mz E_7\oplus \mz E_8 ,\\
K_X&=&-4E+2(E_1+E_2+E_3+E_4+E_5+E_6+E_7+E_8),
\een
where $E$ denotes the total transform of the divisor class of
the plane in $\pth$ and $E_i$ denotes the total transform of the 
exceptional divisor generated by blow-up at $P_i$.
We define $\delta$ as 
\be \delta:=-\frac{1}{2}K_X \ee
for the following argument.

We consider the group (written as ${\rm Gr}(\{X\})$) 
of birational transformations on 
the family $\{X_{P_1,\cdots,P_8}\}$ such that
i)  $\vp:X(4,8)\to X(4,8)$ is an automorphism;
here, we denote $\vp(\{P_1,\cdots,P_8\})$ by $\{P_1',\cdots,P_8'\}$
ii) for any $\{P_1,\cdots,P_8\} \in X(4,8)$ the map
$\vp:X_{P_1,\cdots,P_8}\to X_{P_1',\cdots,P_8'}$  is a birational
map preserving the inner product of $\pic(X)$ which is defined by
using the intersection numbers as
$(D,D'):=D\cdot D'\cdot (-\frac{1}{2}K_X)$ for $D,D'\in \pic(X)$,
where for $D,D',D'' \in \pic(X)$, $D\cdot D'\cdot D''$ 
denotes the intersection number
$\pic(X) \times \pic(X) \times \pic(X) \to \mz$.

\begin{thm}\label{e71}
${\rm Gr}(\{X\})$ is the affine Weyl group of type $E_7^{(1)}$.
\end{thm}

Before the proof we describe some formulae for the intersection 
numbers. For elements in $\pic(X)$ the intersection numbers are
given by
\be &&E\cdot E\cdot E=1, ~~  E\cdot E \cdot E_i=0,  \nonumber\\
   && E \cdot E_i \cdot E_i=0, ~~
    E_i\cdot E_i \cdot E_i=1, \nn \\
    && E_i \cdot E_j\cdot D=0~~ (i\neq j, \forall D \in \pic(X))
\ee
and their linear combinations.
Hence the inner product is given by 
\be (E,E)=2, &  (E,E_i)=0, &(E_i,E_j) = -\delta_{i,j}\ee
and their linear combinations.

\begin{lemma}
If a birational map $\vp:X_{P_1,\cdots,P_8}\to X_{P_1',\cdots,P_8'}$
preserves the inner product, then
$\vp_*: \pic(X_{P_1,\cdots,P_8}) \to \pic(X_{P_1',\cdots,P_8'})$
is an isomorphism of lattices.
\end{lemma}

\bp
Note that $\vp_*$ $(=(\vp^{-1})^*)$ can be considered to be a linear 
transformation to itself.
Assume that there exists a nonzero divisor 
$D \in \pic(X_{P_1,\cdots,P_8})$ such that $\vp_*(D)=0$. 
Since $D$ is nonzero, there exists a divisor 
$D' \in \pic(X_{P_1,\cdots,P_8})$ such that
$(D,D')\neq 0$. On the other hand, $\vp_*(D)=0$ and hence 
$(\vp_*(D),\vp_*(D'))=0$, which contradicts the assumption that $\vp$ 
preserves the inner product. \ep\\

By this lemma, each
$\vp_*: \pic(X_{P_1,\cdots,P_8}) \to \pic(X_{P_1',\cdots,P_8'})$
is an isomorphism and preserves \\
a) the inner product;\\
b) the anti-canonical divisor $-K_X$, i.e.
$\vp_*(-K_{X_{P_1,\cdots,P_8}}) = -K_{X_{P_1',\cdots,P_8'}}$, since $\vp$ is birational;\\
c) the effectiveness of divisors.\\

We identify the lattices 
$\pic(X_{P_1,\cdots,P_8})$'s and 
call an automorphism of $\pic(X)$ which preserves
a), b), c) a (3-dimensional) Cremona isometry.\\

\noi {\bf Proof of Theorem \ref{e71}}

As in the 2-dimensional case, the theorem \ref{e71} is proved 
by investigating the group of Cremona isometries 
and realizing the corresponding birational transformations.

Let $\vp\in {\rm Gr}(\{X\})$.
Since $\vp_*$ preserves $K_X$ and the inner product, it also preserves 
the orthogonal complement of $K_X$
$$Q(\al):=<\al_0, \al_1,\cdots, \al_7>_{\mq},$$
where
\be && \al_0=E_1-E_2, ~ \al_1=E_2-E_3,~ \cdots, ~\al_6=E_7-E_8, \nonumber \\
     && \al_7=E-E_1-E_2-E_3-E_4 .\ee

\begin{claim}
The basis $<\al_0, \al_1,\cdots, \al_7>$ of the linear space $K_X^{\bot}$
generates the lattice $<\al_0, \al_1,\cdots, \al_7>_{\mz}$,
i.e.
$$ <\al_0, \al_1,\cdots, \al_7>_{\mq} \cap \pic(X) =  <\al_0, \al_1,\cdots, \al_7>_{\mz} $$
holds.\end{claim}

\bp
We show that the left hand side includes the right hand side.
Let $a_0 \al_0 +\cdots+a_7 \al_7 \in \pic(X)$. 
Since the coefficient of $E_8$ is an integer, $a_6$ is also an integer.
Since the coefficient of $E_7$ is an integer, $-a_5+a_6$ is also an integer, 
and so is $a_5$. Along the same line 
all $a_i$'s are integers. \ep\\

From this claim, $\vp_*$ is an automorphism of the sub-lattice 
$Q(\al)$ and preserves the inner product.
The matrix defined by using the inner product as
\be (c_{i,j})_{i,j}&:=& 2\frac{(\al_i,\al_j)}{(\al_i,\al_i)} \label{cartan}
\ee
is the affine Cartan matrix of type $E_7^{(1)}$.
We denote the affine Weyl group generated by 
\be &&r_j(\al):=r_{\al_j}(\al)=\al - 2\frac{(\al_j,\al)}{(\al_j,\al_j)}\al_j
\quad (\al \in Q(\al) ) \label{rial} \ee
($i=0,1,\cdots,7$) by $W(E_7^{(1)})$.
\begin{figure}[ht]
\hspace{2cm}
\begin{picture}(500,60)
\put(50,20){\line(1,0){20}}
\put(75,20){\line(1,0){20}}
\put(100,20){\line(1,0){20}}
\put(125,20){\line(1,0){20}}
\put(150,20){\line(1,0){20}}
\put(175,20){\line(1,0){20}}
\put(122.5,22.5){\line(0,1){20}}
\put(47.5,20){\circle{5}}
\put(72.5,20){\circle{5}}
\put(97.5,20){\circle{5}}
\put(122.5,20){\circle{5}}
\put(147.5,20){\circle{5}}
\put(172.5,20){\circle{5}}
\put(197.5,20){\circle{5}}
\put(122.5,45){\circle{5}}
\put(47.5,30){\makebox(0,0){}}
\put(197.5,30){\makebox(0,0){}}
\put(47.5,10){\makebox(0,0){$\al_{0}$}}
\put(72.5,10){\makebox(0,0){$\al_{1}$}}
\put(97.5,10){\makebox(0,0){$\al_{2}$}}
\put(122.5,10){\makebox(0,0){$\al_{3}$}}
\put(147.5,10){\makebox(0,0){$\al_{4}$}}
\put(172.5,10){\makebox(0,0){$\al_{5}$}}
\put(197.5,10){\makebox(0,0){$\al_{6}$}}
\put(140,45){\makebox(0,0){$\al_{7}$}}
\end{picture}
\caption[]{the Dynkin diagram of type $E_7^{(1)}$}
\end{figure}
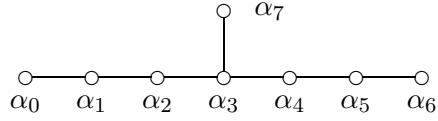
By the following proposition by Kac we have the fact that 
the group of isometries of $Q(\al)$ each of which preserves
the inner product is $\pm {\rm Aut}({\rm Dynkin}) 
\ltimes W(E_7^{(1)})$ (which is written as $\pm \widetilde{W}$). 

\begin{prop}{\bf (\cite{kac} \S 5.10)}
If the generalized Cartan Matrix $c_{ij}$ is a symmetric matrix
of finite, affine, or hyperbolic type, then the group of all automorphisms of
$Q(\al)$ preserving the bilinear form is $\pm \widetilde{W}$.
\end{prop}

Note that 
\be \delta&=&-\frac{1}{2}K_X= \al_0+2\al_1+3\al_2+4\al_3+3\al_4+
2\al_5+\al_6+2\al_7 \ee
is preserved by $W(E_7^{(1)})$ and the automorphism of the Dynkin diagram.
In our case, since $\vp_*$ preserves the anti-canonical divisor $-K_X$,
each element of $-{\rm Aut}({\rm Dynkin}) \ltimes W(E_7^{(1)})$
is not a Cremona isometry.

\begin{claim}
The automorphism of the Dynkin diagram is not a Cremona isometry.
\end{claim}

\bp
There exists only one automorphism of the Dynkin diagram, 
which is the involution exchanging 
$\al_0,\al_1,\al_2$ and $\al_6,\al_5,\al_4$ respectively.

Denoting the actions of these involutions on $\pic(X)$
by $D \mapsto \ol{D}$, we have
\ben \ol{E} &=&  E+3E_1-E_2-E_3-E_4-E_5-E_6-E_7-E_8+4\ol{E_8}\\
    \ol{E_1} &=& E_1-E_8+\ol{E_8}\\
    \ol{E_2} &=& E_1-E_7+\ol{E_8}\\
            &\vdots&\\
    \ol{E_7} &=& E_1-E_2+\ol{E_8}.
\een    
Set $\ol{E_8}=eE+e_1E_1+\cdots+e_8E_8$.
From
\ben (\ol{E},\ol{E_8})&=&(E,E_8)=2e-3e_1+e_2+e_3+e_4+e_5+e_6+e_7+e_8-4=0\\
     (\ol{E_1},\ol{E_8})&=&(E_1,E_8)=-e_1+e_8-1=0\\
        &\vdots&\\
     (\ol{E_7},\ol{E_8})&=&(E_7,E_8)=-e_1+e_2-1=0,
\een
we have $e=-2e_1-3/2$, which contradicts the assumption 
that $e$ is an integer. \ep\\

Now we have the fact that the actions of Cremona isometries on $Q(\al)$
are included by $W(E_7^{(1)})$. 

\begin{claim}
The action of $W(E_7^{(1)})$ are uniquely extended onto $\pic(X)$ as
\be r_j(D)=D - 2\frac{(\al_j,D)}{(\al_j,\al_j)}\al_j
\quad (D \in \pic(X) ). \ee
\end{claim}

\bp
Let $s$ and $s'$ be Cremona isometries such that the action of $s$ is
identical to that of $s'$ on $Q(\al)$.
We show $s'\circ s^{-1}= \mbox{Identity}$ on $\pic(X)$.
Since $\{E_1,\al_0,\al_1,\cdots,\al_7\}$ is a basis of $\pic(X)$, we can set
$$s'\circ s^{-1}(E_1)=e_1 E1 + a_0 \al_0 + a_1 \al_1+\cdots+a_7 \al_7.$$
From $(s'\circ s^{-1}(E_1),\delta)=(E_1,\delta)=1$ and
$(\al_i,\delta)=0$ (for $\forall i$), we have $e_1=1$.
Since 
$$(s'\circ s^{-1}(E_1), \al_i) = (E_1,\al_i)
\Longleftrightarrow 
(s'\circ s^{-1}(E_1)- E_1, \al_i) =0$$
holds,
$$ - (c_{i,j})_{0\leq i,j\leq 7} {\bf a}=0$$
holds, where $(c_{i,j})_{0\leq i,j\leq 7}$ is the Cartan matrix of type 
$E_7^{(1)}$ (\ref{cartan}).
Hence we have
$s'\circ s^{-1}(E_1)=E_1+z \delta$ $(z \in \mz)$.
Finally, from
$(s'\circ s^{-1}(E_1),s'\circ s^{-1}(E_1))
= -1+2z=-1$, we have $z=0$.
Hence it has been shown that $s'\circ s^{-1}$ 
does not change the basis of $\pic(X)$ and therefore
$s'\circ s^{-1}= \mbox{Identity}$. \ep\\

By this claim, the actions of simple reflections on 
$\pic(X)$ are given by
\ben r_i:&& E_{i+1}\mapsto E_{i+2} \nn \\
        && E_{i+2}\mapsto E_{i+1}  \quad (\mbox{for } 0 \leq i \leq 6) 
                         \nn \\
    r_7:&& E \mapsto 3E-2E_1-2E_2-2E_3-2E_4 \nn \\
        && E_k \mapsto E-E_1-E_2-E_3-E_4 +E_k  
                  \quad   (\mbox{for } 1 \leq k \leq 4),
\een
where preserved elements are omitted.
Finally, we have the claim of the theorem by the following claim.

\begin{claim}\label{real}
The action of each element of $W(E_7^{(1)})$ on $\pic(X)$ is uniquely realized
as an element of ${\rm Gr}(\{X\})$.
\end{claim}

\proof
It is enough to show for the simple reflections $r_i$'s.\\
i) in the case of $0 \leq i \leq 6$. 
Since $E \mapsto E$, the action on $\pth$ is linear. This is nothing but 
the projective transformation $PGL(4)$.
Hence, without loss of generality, we can assume that the action is
the identity.
Since $r_i$ exchanges $E_{i+1}$ and $E_{i+2}$, we have
\be  r_i: && ((P_1,\cdots, P_{i+1},P_{i+2},\cdots,P_8); {\bf x}) 
\in X(4.8)\times \pth \nonumber \\
&& \mapsto ((P_1,\cdots, P_{i+2}, P_{i+1},\cdots,P_8); {\bf x}). 
\ee
Here, we have described the action 
$r_i:X_{P_1,\dots,P_8}\to X_{P_1',\dots,P_8'}$
in terms of the coordinate ${\bf x}$ of $\pth$.\\
ii) in the case of $r_7$.
Using $PGL(4)$, we may assume
\be P_1=P_1'=(1:0:0:0),\cdots,P_4=P_4'=(0:0:0:1) \label{nor1} \ee
without loss of generality.
Since $E \mapsto 3E-2E_1-2E_2-2E_3-2E_4$,
the action on $\pth$: $(x:y:z:w)\mapsto(x':y':z':w')$ is in the 3rd degree.
Moreover, from $E_1 \leftrightarrow E-E_2-E_3-E_4$, we have
$(1:0:0:0)\leftrightarrow x'=0$. Since similar facts hold for the case
of $E_2,E_3$ and $E_4$, considering the degree with respect to 
$x,y,z,w$, we have $(x':y':z':w')=(ayzw:bzwx:cwxy:dxyz)$, 
where $a,b,c,d \in \mc^{\times}$. Furthermore, 
we can normalize it as $a=b=c=d=1$ preserving (\ref{nor1}).
This is nothing but the standard Cremona transformation of $\pth$.
Hence the action of $r_7$ on $\{X\}$ is given by the composition of
maps:
\be  &&((P_1,P_2,P_3,P_4),(P_5,P_6,P_7,P_8); {\bf x}) 
\in X(4.8)\times \pth \nonumber \\
& \mapsto& 
\left( \left( \ba{llll} 1&0&0&0\\ 0&1&0&0\\ 0&0&1&0\\ 0&0&0&1\ea \right),
  P_{1234}^{-1} ( P_5,P_6,P_7,P_8); 
  P_{1234}^{-1}\left(\ba{l}x\\y\\z\\w\ea\right) \right) \nonumber\\ 
&& :=((P_1',\cdots,P_8');{\bf x}') \nonumber \\ 
& \mapsto& 
\left( \left( \ba{llll} 1&0&0&0\\ 0&1&0&0\\ 0&0&1&0\\ 0&0&0&1\ea \right),
  \left( \ba{lll} 
  x_{2,5}'x_{3,5}'x_{4,5}'&\cdots&x_{2,8}'x_{3,8}'x_{4,8}'\\ 
  x_{3,5}'x_{4,5}'x_{1,5}'&\cdots&x_{3,8}'x_{4,8}'x_{1,8}'\\
  x_{4,5}'x_{1,5}'x_{2,5}'&\cdots&x_{4,8}'x_{1,8}'x_{2,8}'\\ 
  x_{1,5}'x_{2,5}'x_{3,5}'&\cdots&x_{1,8}'x_{2,8}'x_{3,8}'
  \ea \right); 
  \left(\ba{l}y'z'x'\\z'w'x'\\w'x'y'\\x'y'z'\ea\right) \right),   
\ee
where $P_{1234}$ denotes the square matrix $(P_1,P_2,P_3,P_4)$. \ep\\

\ep {\bf Proof of Theorem \ref{e71}}

\section{Normalization of the configuration space}

In this section we discuss normalization of the configuration space
$X(4,8)$ by $PGL(4)$, i.e. how to choose representative elements.
Notice that without good normalization it is difficult to see 
the concrete action of the group and properties of associated 
dynamical systems.  
For example, although $X(4,8)$ is easily normalized as
\be 
&& \left(\ba{cccccccc}
1&0&0&0&1&1&1&1\\
0&1&0&0&1&y_6&y_7&y_8\\
0&0&1&0&1&z_6&z_7&z_8\\
0&0&0&1&1&w_6&w_7&w_8\ea\right),
\ee
the action of the Weyl group on this coordinate becomes
complicated. 
We normalize it in terms of elliptic curves.

Note the following lemma.

\begin{lemma}
Let $P_1,P_2,\cdots,P_8$ be 8 points in $\pth$.
Quadratic surfaces passing through $P_1,P_2,\cdots,P_8$ form 
at least one parameter family.
\end{lemma}

\bp
It is clear from the fact that
quadratic surfaces passing through 
$P_1,P_2,\cdots,P_8$ and  arbitrary point $P_9$ in $\pth$
is given by the equation
\ben
\left| \ba{cccccccccc}
x^2&y^2&z^2&w^2&xy&yz&zw&wx&xz&yw\\
x_1^2&y_1^2&z_1^2&w_1^2&x_1y_1&y_1z_1&z_1w_1&w_1x_1&x_1z_1&y_1w_1\\
&&&&\vdots&&&&&\\
x_9^2&y_9^2&z_9^2&w_9^2&x_9y_9&y_9z_9&z_9w_9&w_9x_9&x_9z_9&y_9w_9
\ea \right|&=&0,
\een
where $P_i=(x_i:y_i:z_i:w_i)$
(if the left hand side is identically zero, exchange one of the points
for a generic point). \ep\\

The pencil of quadratic surfaces passing through $P_1,P_2,\cdots,P_8$
can be written as 
\be {\bf x}^t (\al A+\beta B){\bf x}=0&& ({\bf x}\in \pth) \label{ab} \ee
by $4 \times 4$ complex symmetric matrices
$A,B$ and $(\al:\beta)\in \pone$.
Normalizing (\ref{ab}) by $PGL(4)$ (cf. \cite{yoshida}), 
we have the next theorem, which
provides a normalization of $X(4,8)$.
The proof of this theorem will be given in the last part of this section.

\begin{thm} \label{norp}
Each element of $X(4,8)$ can be normalized
so that $P_1,P_2,\cdots,P_8$ are on the intersection curve(s) 
of one of the following 3 type pencils of quadratic surfaces.\\
{\rm (i)}
\ben  (E) & x^2-zw=0\\
      (F) & y^2-4xw+g_2xz+g_3z^2=0
\een 
{\rm (ii)}
\ben   && xy-zw=0\\
       && xw-z^2=0
\een 
{\rm (iii)}
\ben   && xy-zw=0\\
       && 4x^2-2zw+w^2=0  ~.
\een 
Moreover,\\

\noi {\rm (i-1)} in the case where the intersection curve is non-singular
$(\Delta=27g_3^2-g_2^3\neq0)$,
$P_i$ can be parameterized as
\be \label{curvei1}
P_i=(\wp(u_i):\wp'(u_i):1:\wp^2(u_i)), \qquad u_i\in \mc/(\mz+ \mz\tau),
\ee
where $\wp(u)$ is Wierstrass $\wp$ function with the periods $(1,\tau)$.\\
{\rm (i-2)} in the case where $\Delta=0$ and $g_2\neq 0$,
the intersection curve can be renormalized to the intersection curve of
\ben  x^2-zw&=&0\\
      y^2-4w(x+h z)&=&0
\een 
and $P_i$ can be parameterized as
\be \label{curvei2}
P_i&=&\left( \frac{4 h u_i}{(1-u_i)^2} : 
\frac{-8 h^{3/2} u_i(1+u_i)}{(1-u_i)^3} :1:
\frac{(4 h u_i)^2}{(1-u_i)^4} \right), \qquad u_i\in \pone \setminus 
\{0,\infty \} ~.
\ee
{\rm (i-3)} in the case where $g_2=g_3=0$,
the inter section curve can be renormalized to the intersection curve of
\ben  x^2-zw&=&0\\
      y^2-4xw&=&0
\een 
and $P_i$ can be parameterized as
\be \label{curvei3}
P_i&=&\left( u_i^{-2} : 
-2u_i^{-3} :1: u_i^{-4} \right),  \qquad u_i\in \pone \setminus \{\infty \}~.
\ee
{\rm (ii)} the intersection consists of 2 curves 
$\{(0:s:0:t)~|~s:t\in \pone\}$ and
$\{(s^3:t^3:s^2t:st^2)~|~s:t\in \pone\}$.\\
{\rm (iii)} the intersection consists of 2 curves
$\{(0:s:0:t)~|~s:t\in \pone\}$ and
$\{(2st^2:s(s^2+4t^2):t(s^2+4t^2):2s^2t)~|~s:t
\in \pone\}$.
\end{thm}

\remark
The parameterizations of (i-2) and (i-3) of Theorem ~\ref{norp}
are chosen so that the period map becomes simple.\\

From the proof in case (i-1) of this theorem 
we have the next corollary.

\begin{cor}
If the intersection curve of surfaces passing through
$P_1,P_2,\cdots,P_8$ is non singular,
$X(4,8)$ can be normalized as $P_1,P_2,\cdots,P_8$
are on the intersection curve of 2 surfaces
\be  \theta_{00}^2x^2-\theta_{01}^2y^2-\theta_{10}^2z^2&=&0  \nn \\
     \theta_{10}^2y^2-\theta_{01}^2z^2-\theta_{00}^2w^2&=&0 ,
\ee 
where we write $\theta_{ij}:=\theta_{ij}(0)$ for 
the theta function $\theta_{ij}(u)$ with the fundamental
periods $(1,\tau)$ and therefore $P_i$ can be parameterized as 
\be
P_i=(\theta_{00}(2u_i):\theta_{01}(2u_i):
\theta_{10}(2u_i):\theta_{11}(2u_i)).
\ee
\end{cor}

\begin{cor}\label{noru}
$X(4,8)$ restricted to case
${\rm (i-1)},{\rm (i-2)}$ or ${\rm (i-3)}$ of Theorem \ref{norp}
is isomorphic to\\
{\rm (i-1)}
\be &&\{(u_1,\cdots,u_8) \in (\mc/(\mz+\mz\tau))^8,
~\tau \in {\mathbb H}/SL(2,\mz)~|~u_i+u_j+u_k+u_l \neq 0\}, \ee
where $1\leq i,j,k,l \leq 8$ are different each other and 
${\mathbb H}$ is the upper half of the complex plane;\\
{\rm (i-2)}
\be &&\hspace{-1cm}\{(u_1,\cdots,u_8)\in 
(\pone \setminus \{0,\infty \})^8~|~u_i u_j u_k u_l \neq 1 \};\ee
{\rm (i-3)}
\be &&\hspace{-1cm}\{(u_1,\cdots,u_8)\in 
(\pone \setminus \{\infty \})^8~|~u_i+u_j+u_k+u_l \neq 0 \}. \ee
\end{cor}

\bp It is enough to show that 
$P_i,P_j,P_k,P_l$ are on the same plane if and only if 
$u_i+u_j+u_k+u_l = 0$ holds
($u_iu_ju_ku_l=1$ holds in case (i-2)).
Notice that$P_i,P_j,P_k,P_l$ are on the same plane
if and only if 
\be \left| \ba{cccc} x_i&y_i&z_i&w_i\\ x_j&y_j&z_j&w_j\\x_k&y_k&z_k&w_k\\
x_l&y_l&z_l&w_l \ea \right| =0 \label{plane} \ee
holds.\\
$\cdot$ In case (i-1). When (\ref{plane}) is considered to be 
a rational function of $u_i$, the origin is the unique pole of order 4.
By Abel's Theorem the sum of zero points are $0$.
Here, $u_i=u_j,u_k,u_l$ are zero points and hence the other zero point
is $u_i=-u_j-u_k-u_l$.\\
$\cdot$ In case (i-2) or (i-3). When (\ref{plane}) is considered to be 
a rational function of $u_i$, $u_i=1$ (the origin in case (i-3)) 
is the unique pole 
of order 4 and therefore there are 4 zero points. 
It is easily shown that $u_i=u_j,u_k,u_l,(u_ju_ku_l)^{-1}$, 
($u_i=u_j,u_k,u_l,-u_j-u_k-u_l$ in case (i-3)) are those 4 points 
by substitution.
\ep\\

A similar argument leads the following theorem 
concerning the dimension of linear system 
$|-\frac{1}{2}K_X|$ of rational variety $X=X_{P_1,\cdots,P_8}$.

\begin{thm}\label{ell}
In case (i) of theorem \ref{norp},
$\dim(|-\frac{1}{2}K_X|)$ is 2 or more if and only if, 
in case {\rm (i-1)} or {\rm (i-2)}
\be u_1+u_2+\cdots+u_8&=&0\ee
holds
and 
in case {\rm (i-1)}
\be u_1u_2\cdots u_8&=&1\ee
holds.
\end{thm}

Notice that if the equivalent conditions of 
Theorem \ref{ell} is satisfied, $X$ is an elliptic variety.\\

\bp
Case (i-1). \\
($\Rightarrow$). Since $\dim(|-\frac{1}{2}K_X|)\geq 2$,
there exists a surface $D\in |-\frac{1}{2}K_X|$ which 
does not include the intersection curve of (i-1).
Let $P_9$ and $P_{10}$ be generic points on $D$.
$D$ is described as
\be \label{minus7}
\left| \ba{cccccccccc}
x^2&y^2&z^2&w^2&xy&yz&zw&wx&xz&yw\\
x_1^2&y_1^2&z_1^2&w_1^2&x_1y_1&y_1z_1&z_1w_1&w_1x_1&x_1z_1&y_1w_1\\
&&&&\vdots&&&&&\\
x_7^2&y_7^2&z_7^2&w_7^2&x_7y_7&y_7z_7&z_7w_7&w_7x_7&x_7z_7&y_7w_7\\
x_9^2&y_9^2&z_9^2&w_9^2&x_9y_9&y_9z_9&z_9w_9&w_9x_9&x_9z_9&y_9w_9\\
x_{10}^2&y_{10}^2&z_{10}^2&w_{10}^2&x_{10}y_{10}&y_{10}z_{10}&
z_{10}w_{10}&w_{10}x_{10}&x_{10}z_{10}&y_{10}w_{10}
\ea \right|&=&0 . 
\ee
By B\'ezout's theorem
the intersection of $D$ and the curve of (i-1) is the 8 points 
$P_1,P_2,\cdots,P_8$, which are given by the zero points of (\ref{minus7})
with $(x:y:z:w)=(\wp(u):\wp(u)':1:\wp(u)^2)$.
The left hand side of the equation is an elliptic functions of $u$
and has the unique pole $u=0$ of order $8$.
Since the 8 points $u=u_1,u_2,\cdots,u_8$ are zero points, by Abel's theorem we
have $u_1+u_2+\cdots+u_8=0$.\\
($\Leftarrow$)
Assume $u_1+u_2+\cdots+u_8=0$. 
Considering the zero points of the same elliptic function,
it is shown that the intersection of a generic quadratic 
surface $D$ passing through 7 points $P_1,P_2,\cdots,P_7$ and 
the curve of (i-1) is the 8 points
$P_1,P_2,\cdots,P_8$. Hence $D$ is an element of $|-\frac{1}{2}K_X|$.
The dimension to choose such $D$ is 2 or more.\\
In case (i-2) or (i-3), it is enough to change the argument about
Abel's theorem as the proof of Corollary \ref{noru}.
\ep\\

\noi{\bf Proof of Theorem \ref{norp}}

Since the later part is easy, we show only the former part.

We consider normalization of (\ref{ab}) by $PGL(4)$.
Note that $P^{-1}$ in $PGL(4)$ acts on the pencil as
$${\bf x}^t (\al A+\beta B){\bf x}=0 \to 
{\bf x}^t (\al P^tAP+\beta P^tBP){\bf x}=0.$$
Note also the following facts.\\

\noi Fact 1.
Any complex symmetric matrix can be diagonalized to
the form ${\rm diag}(1,\cdots,1,0,\cdots,0)$ by $PGL$.

\noi Fact 2.
The $n\times n$ identity matrix is not changed by
orthonormal matrices.

\noi Fact 3.
Since ``two complex symmetric matrices are similar if and only if 
they are similar via a complex orthonormal similarity,''
if two matrices have the same Jordan normal form, 
they are mapped each other by some complex orthonormal matrix.
(pp.212 in \cite{hj}).\\

Assume that there exists $(s:t)$ such that $\rank (sA+tB)=2$,
by Fact 1 the matrix is normalized to ${\rm diag}(1,1,0,0)$
and the defining equation can be factorized.
Hence, 4 or more points in the 8 points on ${\bf x}^t (\al A+\beta B){\bf x}=0 $
are on the same plane, which contradicts the assumption of 
the configuration space.
So we have $\rank(sA+tB)\geq 3$ for all $(s:t)\in \pone$.

\begin{lemma}
If $\rank(sA+tB)\geq 3$ for all $(s:t)\in \pone$,
there exists $(s:t)\in \pone$ such that $\rank(sE+tF)=4$.
\end{lemma}

\bp
Without loss of generality we can assume $A= {\rm diag}(1,1,1,0)$.
We normalize $B$ by $3 \times 3$ matrix.
Since the sizes of Jordan blocks of the $3\times 3$ submatrix of $B$
should be $(1,1,1), (2,1)$ or $(3)$,
by Fact 3, we can set $B$ as
\ben  &&\left( \ba{cccc} a&0&0&e\\ 0&b&0&f\\ 0&0&c&g\\ e&f&g&d \ea \right),
\quad
\left( \ba{cccc} a+1&\sqrt{-1}&0&e\\ \sqrt{-1}&a-1&0&f\\ 0&0&b&g\\ e&f&g&d 
\ea \right), \quad
\left( \ba{cccc} a&1&\sqrt{-1}&e\\ 1&a&0&f\\ \sqrt{-1}&0&a&g\\ e&f&g&d \ea \right).\een
Assume $\rank(sA+tB)= 3$ for all $(s,t)\in \pone$.
In the first case, we have $d=e=f=g=0$ and therefore 
the rank becomes $2$ or less 
for some $t$, which is a contradiction.
Along the same line, in the second or the third case
it can be shown that the rank becomes 2 or less, 
which is a contradiction. 
Hence there exists $(s:t)\in \pone$ such that $\rank(sA+tB)= 4$.\ep\\

From the above lemma, we may assume
$A={\rm Identity}$ and $\rank{B}=3$. 
The Jordan normal form of $B$ has the following 5 possibilities:
\ben 
{\rm (i-1)}&&\left(\ba{cccc} a&0&0&0\\ 0&b&0&0\\ 0&0&c&0\\ 0&0&0&0 \ea 
\right), \\
{\rm (i-2)}&&\left(\ba{cccc} a&1&0&0\\ 0&a&0&0\\ 0&0&b&0\\ 0&0&0&0 \ea\right)
\sim \left(\ba{cccc} 0&0&0&0\\ 0&a+1&\sqrt{-1}&0\\ 0&\sqrt{-1}&a-1&0\\ 0&0&0&1 \ea\right),\\
{\rm (i-3)}&&\left(\ba{cccc} a&1&0&0\\ 0&a&1&0\\ 0&0&a&0\\ 0&0&0&0 \ea \right)
\sim \left(\ba{cccc} 0&0&0&0\\ 0&1&1&0\\ 0&0&1&1\\ 0&0&0&1 \ea\right),\\
{\rm (ii)}&&\left(\ba{cccc} 0&1&0&0\\ 0&0&1&0\\ 0&0&0&1\\ 0&0&0&0 \ea\right) 
\sim \left(\ba{cccc} 0&0&1&\sqrt{-1}\\ 0&0&\sqrt{-1}&-1\\ 1&\sqrt{-1}&-1&\sqrt{-1}\\ \sqrt{-1}&-1&\sqrt{-1}&1 \ea\right),\\
{\rm (iii)}&&\left(\ba{cccc} 0&1&0&0\\ 0&0&0&0\\ 0&0&a&1\\ 0&0&0&a \ea\right) 
\sim \left(\ba{cccc} 1&\sqrt{-1}&0&0\\ \sqrt{-1}&-1&0&0\\ 0&0&2&\sqrt{-1}\\ 0&0&\sqrt{-1}&0 \ea\right), 
\een
where the right hand side matrices are
similar to $B$ except the proportional constants.
We replace $B$ by the right hand side matrices.

\noi $\cdot$ Case (i-1). 
Since $\rank(sA+tB)\geq 3$, $a,b,c$ are not zero and different each other.
By replacing the basis of pencil ${\bf x}^t (\al A+\beta B){\bf x}=0$,
we may assume $A={\rm diag}(0,a,b,c)$ and $B={\rm diag}(d,e,f,0)$.
Moreover, by the actions of diagonal matrices,
we can set $A={\rm diag}(0,1,1,1)$ and $B={\rm diag}(a,a,b,0)$.
Finally, by considering ${\rm constant}\times B$, we can set 
$A={\rm diag}(0,1,1,1), B={\rm diag}(1,1,a,0)$ ($a\neq 0,1$).
On the other hand if $\Delta=g_2^3-27g_3^2$ is not zero, 
using $e_1=\wp(w_1/2),e_2=\wp(w_2/2),e_3=\wp((w_1+w_2)/2)$,
$(F)$ can be written as 
$$y^2-4xw + 4(e_1 + e_2 + e_3)x^2 - 4(e_1 e_2 + e_1 e_3 + e_2 e_3)x z  
  + 4e_1 e_2 e_3 z^2=0 .$$
The matrices
\ben E&=& \left(\ba{cccc} 1&0&0&0\\ 0&0&0&0\\ 
0&0&0&-\frac{1}{2}\\ 0&0&-\frac{1}{2}&0 \ea \right),\\
 F&=& \left(\ba{cccc} 4(e_1 + e_2 + e_3)&0&-2(e_1 e_2 + e_1 e_3 + e_2 e_3)&-2\\ 0&1&0&0\\ 
-2(e_1 e_2 + e_1 e_3 + e_2 e_3)&0&4e_1 e_2 e_3&0\\ 
-2&0&0&0 \ea \right) \een
are normalized to
$E'={\rm diag}(0,1,1,1), F'={\rm diag}(1,1,\lambda,0)$ via
\ben P&=& \left(\ba{cccc} 0&1&0&0\\ 1&0&0&0\\ 
0&0&\sqrt{-1}&-1\\ 0&0&\sqrt{-1}&1 \ea \right), \een
where
\be \lambda&=&\frac{e_2-e_3}{e_1-e_3} \nn \\
&=& \frac{3^{1/3}(3 \sqrt{-1} + \sqrt{3})g_2 - (-3 \sqrt{-1} + \sqrt{3})
   (9 g_3 + \sqrt{-3 g_2^3 + 81 g_3^2})^{2/3}}{
   -3^{1/3}(-3 \sqrt{-1} + \sqrt{3})g_2 +
  (3\sqrt{-1} + \sqrt{3})(9g_3 + \sqrt{-3g_2^3 + 81g_3^2})^{2/3}} 
\label{xg2g3} \ee
is the $\lambda$ function. By suitably replacing $P$, $\lambda$ can be
changed to $1/\lambda,1-\lambda,1/(1-\lambda),\lambda/(\lambda-1),
(\lambda-1)/\lambda$
(by using the fact that $E'$ and $F'$ are simultaneously decomposed to the 
eigenspaces, it can also be shown that $\lambda$ cannot be other than these.)
Note that $\lambda$ is invariant under the action 
$(w_1,w_2)\mapsto (sw_1,sw_2)\Leftrightarrow
(g_2,g_3)\mapsto (g_2/s^4,g_3/s^6)$.
We show that there exist corresponding $g_2$ and $g_3$ when
$\lambda \neq 0,1,\infty$
($\Delta=0$ if $\lambda=0,1$ or $\infty$).
Setting
\ben y=\frac{\lambda-1/2-\sqrt{3}\sqrt{-1}/2}{\lambda-1},
\een
we have
$$y=\frac{3 \sqrt{-1}(1/2- \sqrt{3}/2) g_2}{g_2 - 
(3\sqrt{3} g_3 + \sqrt{-g_2^3 + 27 g_3^2})^{(2/3)}}.$$
We show that there exist corresponding $g_2$ and $g_3$ when
$y \neq 1/2+\sqrt{3} \sqrt{-1}/2,1,\infty$.
Setting 
$g_2=a, (3\sqrt{3} g_3 + \sqrt{-g_2^3 + 27 g3^2})^{(2/3)}=3 b^2,$
we have 
$$y=\frac{3 \sqrt{-1}(1/2- \sqrt{3}/2) a}{a - 3 b^2}.$$
Hence, there exist corresponding $a,b\in \mc$ for
arbitrary $y \in \mc$.
Since
$g_2=a,g_3=a^3/(54 b^3)+b^3/2$, 
we can find the corresponding $g_2$ and  $g_3$ when $b \neq 0$.
If $b=0$, we have $g_2=g_3=0$, which is not in the case 
considered here.\\

\noi $\cdot$ Case (i-2).
Set
\ben
&&P_1=\left(\ba{cccc}1&0&0&0\\0&1&0&0\\0&0&2&0\\0&0&0&1\ea \right), \qquad
P_2= \left( \ba{cccc}1&0&0&0\\0& \frac{-\sqrt{-1}}{\sqrt{a}}&\frac{-1}{(a-1)
 \sqrt{a}}&0\\
0&0& \frac{\sqrt{-1} \sqrt{a}}{2(1 -  a)}&0\\0&0&0&\sqrt{\frac{-a}{a-1}} \ea \right),\\
A'&=&P_2^t (P_1 A P_1-P_1 B P_1) P_2 ,\\
B'&=&\frac{1-a}{a} P_2^t P_1 B P_1 P_2 ,
\een
Then $A'={\rm diag}(1,1,1,0)$ and 
the Jordan normal form of $B'$ is
\be \label{jor2}
\left(\ba{cccc}0&0&0&0\\ 0&1&1&0 \\0&0&1&0\\0&0&0&1\ea \right). \ee
On the other hand, if $\Delta=0$, setting 
\ben
&&P_1=\left(\ba{cccc}0&1& 0& 0\\1& 0& 0& 0\\0& 0& \sqrt{-1}& -1\\0& 0& \sqrt{-1}& 1\ea
\right), \qquad
P_2=\left( \ba{cccc} \sqrt{2 \sqrt{3 g_2}}& 0& 0& 0\\0& 1& 0& 0\\
0& 0& 1& 0\\0& 0& 0& 1\ea\right), \\
&& E'=P_2 P_1^t A P_1 P_2,\\
&& F'=-\frac{2}{3} P_2( P_1^t B P_1
\frac{\sqrt{3}}{4 \sqrt{g_2}}-P_1^t A P_1) P_2 ,
\een
we have $E'={\rm diag}(1,1,1,0)$ and (\ref{jor2}) as the Jordan normal
form of $F'$.
Hence it has been shown that the two pencils are equivalent modulo
$PGL(4)$.

\noi $\cdot$ Case (i-3).
If $g_2=g_3=0$,
setting
\ben &&P=\left( \ba{cccc}0& 1& \frac{2 \sqrt{-1}}{\sqrt{5}}&
\frac{2}{\sqrt{5}}\\1& 0& 0& 0\\
0& 0& \frac{\sqrt{-1}}{\sqrt{5}}& \frac{-9}{\sqrt{5}}\\
0& 0& \frac{\sqrt{-1}}{\sqrt{5}}& \frac{1}{\sqrt{5}}\ea \right),\\
&&E'=P^t (A+B) P,  \qquad F'=P^t A P ,\een
we have $E'={\rm Identity}$ and 
\ben \left(\ba{cccc}0&0&0&0\\ 0&1&1&0 \\0&0&1&1\\0&0&0&1\ea\right) \een
as the Jordan normal form of $F'$.
Hence it has been shown that the two pencils are equivalent modulo
$PGL(4)$.

\noi $\cdot$ Case (ii).
Setting
\ben P&=&\left(\ba{cccc}1& 1/2& 0&0\\-\sqrt{-1}& \sqrt{-1}/2& 0& 0\\0& 0& -1&1/2\\
0& 0&-\sqrt{-1}& -\sqrt{-1}/2\ea \right), \\
&& A'=P^t A P, \\
&& B'=P^t B P ,\een
we have
\ben & A'= \left(\ba{cccc} 0& 1& 0& 0\\1& 0& 0& 0\\
0& 0& 0& -1\\0& 0& -1& 0 \ea\right),
&  B'= \left(\ba{cccc} 0& 0& 0& 2\\0& 0& 0& 0\\0& 0& -4& 0\\2& 0& 0& 0
\ea\right).
\een

\noi $\cdot$ Case (iii).
Setting $P,A'$ and $B'$ as in case (ii),
we have
\ben &A'= \left(\ba{cccc} 0& 1& 0& 0\\1& 0& 0& 0\\0& 0& 0& 
-1\\0& 0& -1& 0 \ea\right),
 & B'= \left(\ba{cccc} 4& 0& 0& 0\\0& 0& 0& 0\\
               0& 0& 0& -1\\0& 0&-1& 1\ea\right).
\een
\ep \quad {\bf Proof of Theorem \ref{norp}} \\


\section{Period map and the action of the Weyl group}

In this section we describe
the action of $W(E_7^{(1)})$ in case (i)
of Theorem \ref{norp} in normalized coordinates. 
For this purpose, it is enough to
normalize the simple reflections obtained in Claim \ref{real},
but it is not easy calculus. Thus, first, we define a linear map $\chi_X$ from
the lattice $Q(\al)$ to $\mc$
(which has ambiguity corresponding to the periods 
as discussed later). Next, we compute the action
by using the fact that $\chi$ is invariant under the action of
$W(E_7^{(1)})$, i.e. 
$\chi_X(\al)=\chi_{w(X)}(w(\al))$ for $\alpha \in Q(\al)$ and 
$w \in W(E_7^{(1)})$.
This method is an analogue of that of the period map
essentially introduced by Looijenga for surfaces.\\

\noi {\bf Period map and the action on 
the intersection curve}\\

In the following, we shall discuss only in case (i-1)
of Theorem \ref{norp}. For case (i-2) and (i-3), we shall
write the results only.
Replace  $x/z,y/z,w/z$ by $x,y,w$. 
Let $D_1,D_2\in -\frac{1}{2}K_X$ the divisors determined by $(E),(F)$
in Theorem \ref{norp}.
We denote the set of piece-wise smooth singular 3-chains 
in $X-D_1-D_2$ by $S(X-D_1-D_2)$.
Using the holomorphic 3-form on $X \setminus (D_1 \cup D_2)$
\be \omega=\frac{c~ dx \wedge dy \wedge dw}{(x^2-w)(y^2-4xw+g_2 x+g_3)}
\ee
(the constant $c \in \mc^{\times}$ is determined later),
we define the map $\chi_X :S(X-D_1-D_2) \to \mc$ by
the paring $\int_{\Gamma}\omega$, ($\Gamma \in S(X-D_1-D_2)$).

Let $C$ denote the elliptic curve $D_1\cap D_2$.
We define a map $Q(\al) \to S(X-D_1-D_2)/H_1(C,\mz)$.
For this purpose, it is enough to define for the basis $\al_i$'s.\\
$\cdot$ In the case of $0\leq i \leq 6$.
We have $\al_i=E_{i+1}-E_{i+2}$. Let $C_i^{re}$ be a real curve
on $C$ from $C\cap E_{i+1}$ (which is expressed as $u=u_{i+1}$ 
in the coordinate $u$) to
$C\cap E_{i+2}$ ($u=u_{i+2}$).
Here, $C$ has ambiguity of $H_1(C,\mz)\simeq \mz+\mz \tau$.
Let $\ve >0$ be a sufficiently small number and
$\Gamma_i \in S(X-D_1-D_2)$ the set of points such that 
$|x^2-w|=\ve$, $|y^2-4xw+g_2x+g_3|=\ve$ and $x$ 
is in the projection of $C_i^{re}$ to the $x$ coordinate.\\
$\cdot$ In the case of $i=7$. We have $\al_7=(E-E_1-E_2-E_3)-E_4$.
Let $C_7^{re}$ be a real curve on $C$
from $C\cap (E-E_1-E_2-E_3)$ ($u=-u_1-u_2-u_3$, cf. Lemma \ref{noru})
to $C\cap E_4$ ($u=u_4$).
Let $\Gamma_7 \in S(X-D_1-D_2)$ be the set of points such that 
$|x^2-w|=\ve$, $|y^2-4xw+g_2x+g_3|=\ve$ and $x$ is
in the projection of $C_7^{re}$ to the $x$ coordinate.

\begin{remark}
As in 2-dimensional case, we can take $\Gamma_i$ from $H_3(X-D_1-D_2,\mz)$.
Let $F_1,F_2$ be divisors such that 
$\al_i$ is written as $\al_i=F_1-F_2$ as above. 
For our purpose, it is enough to add 
singular 3-chains in $F_1$ and $F_2$ to the above $\Gamma_i$. 
Here, the extended part is included by 2-dimensional algebraic
subvariety and hence the effect for the integration is zero.
\end{remark}

By the composition $Q(\al)\to S(X-D_1-D_2)/H_1(C,\mz) \to \mc$,
the map $\chi_X:Q(\al)\to \mc$ is determined modulo the image of $H_1(C,\mz)$.
Let $\pi_x$ denote the projection to the $x$ coordinate.
By residue theorem, we have
\ben \chi_X(\al_i) 
&=& \int_{\Gamma_i}\omega\\
&=&c\int_{ \ba {c}|x^2-w|=\ve \\
            |y^2-4xw+g_2x+g_3|=\ve \\ x\in \pi_x(C_i^{re})\ea }
\frac{dx \wedge dy \wedge dw}{(x^2-w)(y^2-4xw+g_2 x+g_3)}\\
&=&
c'\int_{\ba {c}|y^2-4xw+g_2x+g_3|=\ve \\ 
                                x\in \pi_x(C_i^{re})\ea }
\frac{dx \wedge dy}{(y^2-4x^3+g_2 x+g_3)}\\
&=&
c''\int_{x\in \pi_x(C_i^{re})}
\frac{dx}{y}\\
&=&
c''\int_{{C_i^{re}}^*}du \qquad (x=\wp(u), y=\wp'(u)) \\
&=&
\left\{\ba{ll} c''(u_{i+1}-u_{i+2})  &(0\leq i \leq 6)\\
              c''(-u_1-u_2-u_3-u_4)  &(i=7) \ea
\right. ,
\een
where ${C_i^{re}}^*$ denotes the pullback of $C_i^{re}$ to the space of $u$
and the last result should be considered modulo $c''(\mz+\mz \tau)$.
Since the constant $c\in \mc^{\times}$ has been arbitrary,
we can determine it so that $c''=1$.

By further blow-up along lines, 
the simple reflection $r_i$ can be considered to 
be an exchange of the blow-down structure
of $X=X_{P_1,\cdots,P_8}$ and that 
of $r_i(X)=X_{r_i(P_1,\cdots,P_8)}$, 
i.e. it just changes how to blow-down corresponding to the change
of basis of $\pic(X)$ (cf. Remark \ref{bstr}). 
Let $u=u_0$ denote the intersection point of a effective 
divisor $D$ and the curve $C$.
Since the curve $C$ is preserved by $r_i$ 
(because the modulus of $C$ is not changed),
$r_i(D)$ and $C$ also intersect at $u=u_0$.
Since $D$ is arbitrary,
we have
\ben \chi_X(\al)=\chi_{r_i(X)}(r_i(\al))  & \al \in Q(\al). \een
Considering the composition, 
we have 
\be \chi_X(\al)&=&\chi_{w(X)}(w(\al))  \label{chi}\ee
for all $w\in W(E_7^{(1)})$.

\begin{remark} \label{bstr} This fact can also be considered to be
that $r_i$ is a pseudo-isomorphism from $X$ to 
$r_i(X)$ and determines an exchange of their blow-up points.
\end{remark}

From (\ref{rial}),(\ref{chi}), we have
\be
  &r_i:& (u_{i+1},u_{i+2})\mapsto (u_{i+2},u_{i+1})
                 \qquad (\mbox{for } 0 \leq i \leq 6) \nn \\
 &r_7:&  (u_1,\cdots,u_8) \mapsto
  (u_1-\lambda_1,\cdots,u_4-\lambda_1,
   u_5+\lambda_1,\cdots,u_8+\lambda_1) , \label{riu}
\ee
where $\lambda_1=\frac{1}{2}(u_1+u_2+u_3+u_4)$
(preserved elements are omitted). Moreover, since $r_i$ acts on the elliptic curve $C$ birationally
and therefore it is a translation for the points on $C$ except
$P_j$ and $r_i(P_j)$ $(1\leq j \leq 8)$,
we have
\be r_i:& u=0 \mapsto u=0 & (\mbox{for } 0 \leq i \leq 6) \nn \\
    r_7:& u=0 \mapsto u=\lambda_1 &  .  \label{riu0}
\ee

\noi {\it In case {\rm (i-2)}}

We have
\ben \chi_X(\al_i) 
&=&
c''\int_{{C_i^{re}}^*} \frac{du}{u} \\
&=&
\left\{\ba{ll} c''\log \frac{u_{i+1}}{u_{i+2}}  &(0\leq i \leq 6)\\
              -c''\log(u_1 u_2 u_3 u_4)  &(i=7) \ea
\right.
\een
\ben
 & r_i:& (u_{i+1},u_{i+2})\mapsto (u_{i+2},u_{i+1})
 \qquad (\mbox{for } 0 \leq i \leq 6) \\\\
 &r_7:& (u_1,\cdots,u_8) \mapsto
  (u_1\lambda_1^{-1},\cdots,u_4\lambda_1^{-1},
   u_5\lambda_1,\cdots,u_8\lambda_1) ,
\een
where $\lambda_1=(u_1 u_2 u_3 u_4)^{1/2}$, and we have
\ben r_i:& u=1 \mapsto u=1 & (\mbox{for } 0 \leq i \leq 6)  \\
    r_7:& u=1 \mapsto u=\lambda_1 &  .
\een

\noi {\it In case {\rm (i-3)}}

It is the same with case (i-1).\\

\noi {\bf The action on $\pth$}\\

We investigate the action to generic points
in $X_{P_1,\cdots,P_8}$.\\

\noi {\it The action of $r_i$ {\rm ($0\leq i \leq 6$)}}

Since the simple reflection $r_i$ just exchanges the blow-up points,
it acts on $\pth$ as the identical map
\be  r_i: && {\bf x}\mapsto {\bf x}. 
\ee

\noi {\it The action of $r_7$}

We write $P_i=(f_x(u_i),f_y(u_i),1,f_w(u_i))^t$
by the parametric representation of
$C$: (\ref{curvei1}),(\ref{curvei2}) or 
(\ref{curvei3}).
Let $\ol{*}$ denote the image of $*$ by $r_7$ and
set
$$((P_1,P_2,P_3,P_4),(P_5,P_6,P_7,P_8)) =(A,B).$$
For a matrix $P$, let $1/P$ denote a matrix whose elements
are the reciprocal number of corresponding elements of $P$.
The action of $r_7$ is given by
\ben &&((A,B),{\bf x})\\
& \mapright{A^{-1}}& 
( ( {\rm Id}, A^{-1} B ), A^{-1} {\bf x})\\
& \mapright{\mbox{SCT}}& 
( ({\rm Id},  1/(A^{-1} B) ), 1/(A^{-1} {\bf x}))\\
& \mapright{{\rm diag}\in PGL(4)}& 
{\rm diag}(b_1,b_2,b_3,b_4) 
( ({\rm Id},  1/(A^{-1} B) ), 1/(A^{-1} {\bf x}))\\
& \mapright{\ol{A}}& 
(( \ol{A}, \ol{B} ), \ol{{\bf x}}),
\een
where SCT denotes the standard Cremona transformation.
Setting
${\bf x}'={\rm diag}(b_1,b_2,b_3,b_4) (1/(A^{-1} {\bf x}))$,
we have
$${\bf x}'=\left( \frac{b_1}{\left| {\bf x},P_2,P_3,P_4 \right|},
\frac{-b_2}{\left| {\bf x},P_3,P_4,P_1 \right|},
\frac{b_3}{\left| {\bf x},P_4,P_1,P_2 \right|},
\frac{-b_4}{\left| {\bf x},P_1,P_2,P_3 \right|}
 \right) $$
and
$${\bf x}'= \ol{A}^{-1} \ol{{\bf x}}=
\left( \left| \ol{{\bf x}},\ol{P_2},\ol{P_3},\ol{P_4} \right|,
-\left| \ol{{\bf x}},\ol{P_3},\ol{P_4},\ol{P_1} \right|,
\left| \ol{{\bf x}},\ol{P_4},\ol{P_1},\ol{P_2} \right|,
-\left| \ol{{\bf x}},\ol{P_1},\ol{P_2},\ol{P_3} \right|
 \right). 
$$

On the other hand, from (\ref{riu0}), 
in case (i-1) and case (i-2) we have
\be  \label{xup1} \ol{{\bf p}}&=&
 (f_x(\lambda_1):f_y(\lambda_1):1:f_w(\lambda_1))^t 
\ee
for ${\bf p}=(f_x(0):f_y(0):1:f_w(0))^t=(0:0:0:1)^t$
(in case (i-3) we have $ \ol{{\bf p}} =
(f_x(\lambda_1):f_y(\lambda_1):1:
f_w(\lambda_1))^t$ for
${\bf p}=(f_x(1):f_y(1):1:f_w(1))^t=(0:0:0:1)^t$).
Using these, we can obtain $b_i$ explicitly.
Consequently, we have
\be \label{xup3} \ol{{\bf x}}&=&\left(
\ba{cccc} 
  f_x(\check{u_1})&f_x(\check{u_2})&
  f_x(\check{u_3})&f_x(\check{u_4})\\
    f_y(\check{u_1})&f_y(\check{u_2})&
  f_y(\check{u_3})&f_y(\check{u_4})\\
  1&1&1&1\\
  f_w(\check{u_1})&f_w(\check{u_2})&
  f_w(\check{u_3})&f_w(\check{u_4})
\ea \right)  
\left( \ba{c}
  l_{2,3,4}({\bf x})\\
  -l_{3,4,1}({\bf x})\\
  l_{4,1,2}({\bf x})\\
  -l_{1,2,3}({\bf x})
\ea
\right),
\ee
where
$\check{u_k}:=\ol{u_k}=u_k-\lambda_1$ ($1\leq k \leq 4$)
($\check{u_k}:=\ol{u_k}=u_k \lambda_1^{-1}$ 
($1\leq k \leq 4$) in case (i-2)) and
\be \label{xup4} l_{i,j,k}({\bf x})&=&
\frac{
  \left| 
  \ba{cccc} 
    f_x(\lambda_1)&f_x(\check{u_i})&
    f_x(\check{u_j})&f_x(\check{u_k})\\
    f_y(\lambda_1)&f_y( \check{u_i})&
    f_y(\check{u_j})&f_y(\check{u_k})\\
    1&1&1&1\\
    f_w(\lambda_1)&f_w(\check{u_i})&
    f_w(\check{u_j})&f_w(\check{u_k})
  \ea \right|
  \left| 
  \ba{cccc} 
    0&f_x(u_i)&f_x(u_j)&f_x(u_k)\\
    0&f_y(u_i)&f_y(u_j)&f_y(u_k)\\
    0&1&1&1\\
    1&f_w(u_i)&f_w(u_j)&f_w(u_k)
  \ea \right|
}{
\left| 
  \ba{cccc} 
    x&f_x(u_i)&f_x(u_j)&f_x(u_k)\\
    z&f_y(u_i)&f_y(u_j)&f_y(u_k)\\
    y&1&1&1\\
    w&f_w(u_i)&f_w(u_j)&f_w(u_k)
  \ea \right|
}. \ee


\section{Dynamical systems}

In this section we consider a dynamical system corresponding to
a translation of the Weyl group $W(E_7^{(1)})$.
Note that although one can consider dynamical systems for all translations,
many of them are generated by birational conjugates of
one of them.

Notice that 
$$r_{w(\al_i)}(\beta):= \beta-2\frac{(w(\al_i),\beta)}{(\al_i,\al_i)}w(\al_i)
=w^{-1}\circ r_i \circ w(\beta)$$
holds for $w \in W(E_7^{(1)})$, a simple reflection $\al_i$
and $\beta \in Q(\al)$.
Since the map
\be T&:=&r_{E-E_5-E_6-E_7-E_8}\circ r_{E-E_1-E_2-E_3-E_4}\ee
acts on the root basis as 
\ben
(\al_0,\al_1,\al_2,\al_3,\al_4,\al_5,\al_6,\al_7)
&\mapsto& 
(\al_0,\al_1,\al_2,\al_3+\delta,\al_4,\al_5,\al_6,\al_7-2\delta),
\een
$T$ is a translation. Therefore $T^n$ defines a birational dynamical system on
$X(4,8)\times \pth$. It can also be considered to be 
a dynamical system on $\pth$ with the parameters $u_i$'s (or $P_i$'s).

In case (i-1) or (i-3),
similar to the above section, 
we have the action of $T$ on the parameter space $\{u_i\}$ as
\ben
 && T:  (u_1,\cdots,u_8) \mapsto
  (u_1+\lambda,u_2+\lambda,
  u_3+\lambda,u_4+\lambda,
   u_5-\lambda,\cdots,u_8-\lambda) , 
\een
where $\lambda=\frac{1}{2}\sum_{i=1}^8 u_i$.

Since the explicit action of the transformation on $\pth$
is complicated, we give it using a decomposition.
Although it is enough to compose 
$r_{E-E_5-E_6-E_7-E_8}$ and $r_{E-E_1-E_2-E_3-E_4}$ of course,
here we use the fact $T$ can be written as $T=S^2$ by
\be 
S:=r_{E_4-E_8}\circ r_{E_3-E_7}\circ r_{E_2-E_6}\circ r_{E_1-E_5}\circ
r_{E-E_1-E_2-E_3-E_4} \ee
and describe the action of $S$.
$S$ acts on $\{u_i\}$ as
\be
 && S: (u_1,\cdots,u_8) \mapsto
  (u_5+\lambda_1,u_6+\lambda_1,
  u_7+\lambda_1,u_8+\lambda_1,
   u_1-\lambda_1,\cdots,u_4-\lambda_1)
\ee
and the action on $\pth$ is given by 
(\ref{xup3}) and (\ref{xup4}), where
$\check{u_k}:=u_k-\frac{1}{2}\lambda_1$ ($1\leq k \leq 4$).

Along the same line, in case (i-2) we have
\ben
 && S: (u_1,\cdots,u_8) \mapsto
  (u_5\lambda_1,u_6\lambda_1,
  u_7\lambda_1,u_8\lambda_1,
   u_1\lambda_1^{-1},\cdots,u_4\lambda_1^{-1})
\een
and the action of $S$ on $\pth$ is given by (\ref{xup3}) and (\ref{xup4}),
where $\check{u_k}:=\ol{u_k}=u_k \lambda_1^{-1}$
($1\leq k \leq 4$).


\section{Conservation law}

In this section we prove the next theorem.

\begin{thm}\label{conserve}
In case {\rm (i)} of Theorem \ref{norp}, the action of the Weyl group 
preserves each element of the pencil of quadratic surfaces in 
Theorem \ref{norp}.
\end{thm}

\begin{remark}
The pencil of Theorem \ref{norp} is given by the 
linear system $|-\frac{1}{2}K_X|$ in generic.
By Theorem \ref{conserve}, if the dimension of $|-\frac{1}{2}K_X|$ 
is 2 or more ($\delta=0$ by Theorem \ref{ell}), each fiber
is preserved by translations associated with the Weyl group $W(E_7^{(1)})$,
because
(i) each surface is an elliptic surface (ii) each map is
birational (hence continuous except at the indefinite points)
and preserves the fibration (iii) the modulus of elliptic curve
is preserved (iv) at least the intersection curve of the pencil is preserved.
\end{remark}

Since for every discrete Painlev\'e equation the polynomial degree of
the $n$-th iterate is in the order $n^2$ as 
$n \to \infty$ \cite{takenawa}, 
by Theorem \ref{conserve} the next corollary follows. 

\begin{cor}
Let $\vp$ be a map on $\pth$ associated with a translation
of $W(E_7^{(1)})$. The degree of $\vp^n$ is in the order $n^2$
as $n \to \infty$.
\end{cor}

Since $-\frac{1}{2}K_X$ is preserved and therefore the pencil
itself is preserved, in order 
to prove Theorem \ref{conserve} it is enough to show that
the automorphism of $\pone$ defined by this correspondence is the identity.
In case (i-2) and (i-3) it is easily shown by direct computation.
Moreover, the simple reflection $r_i (0\leq i\leq 6)$  acts on $\pth$
as the identity. Hence it is enough to show for $r_7$ in case (i-1).
For $r_7$, to prove by direct calculation seems to be beyond 
our computational ability. We proof the theorem in this case 
by means of a birational representation of $W(E_8^{(1)})$ on
$\ponet$.

Notice that a smooth quadratic surface is isomorphic to $\ponet$
via the Segr\'e map 
$$ \ponet \ni (x:1,y:1) \mapsto (x:y:1:xy) \in 
\{(x:y:z:w)\in \pth~|~xy-zw=0\} .$$
Here we renormalize the parameter space so that the 8 points are on the 
intersection curve of the pencil spanned by the 2 quadratic surfaces
\be &&\left\{ \ba{ll}
xy-zw=0 & (G) \\
(x+y+z)(4w-\frac{g_3}{\wp^3(2t)}z)=
\left(w+x+y+\frac{g_2}{4\wp^2(2t)}z \right)^2
& (H) \ea \right. , \label{penponet}
\ee
where $t\in (\mc/(\mz+\mz\tau)) \setminus \{0\}$
is an arbitrary extra-parameter.

\begin{remark}\label{modu}
The parameter $\tau=w_2/w_1$ is invariant with respect to $t$.
\end{remark}

By this normalization, $P_i$ can be parameterized as 
$$P_i \left(\frac{\wp(t+u_i)}{\wp(2t)}:\frac{\wp(t-u_i)}{\wp(2t)}:1:
\frac{\wp(t+u_i)\wp(t-u_i)}{\wp^2(2t)}\right).$$
The action of $r_7$ on $({\bf u},t)$ and on $\pth$ are also given by
(\ref{riu}) and (\ref{xup3}) respectively,
where we set 
$$f_x(u)=\frac{\wp(t+u)}{\wp(2t)}, \quad f_y(u)=\frac{\wp(t-u)}{\wp(2t)}, 
\quad f_w(u)=\frac{\wp(t+u)\wp(t-u)}{\wp^2(2t)},$$
$\check{u_k}:=u_k-\lambda_1$ ($1\leq k \leq 4$)
and
\ben l_{i,j,k}({\bf x})&=&
\frac{
  \left| 
  \ba{cccc} 
    f_x(\lambda_1)&f_x(\check{u_i})&
    f_x(\check{u_j})&f_x(\check{u_k})\\
    f_y(\lambda_1)&f_y( \check{u_i})&
    f_y(\check{u_j})&f_y(\check{u_k})\\
    1&1&1&1\\
    f_w(\lambda_1)&f_w(\check{u_i})&
    f_w(\check{u_j})&f_w(\check{u_k})
  \ea \right|
  \left| 
  \ba{cccc}
    1&f_x(u_i)&f_x(u_j)&f_x(u_k)\\
    1&f_y(u_i)&f_y(u_j)&f_y(u_k)\\
    1&1&1&1\\
    1&f_w(u_i)&f_w(u_j)&f_w(u_k)
  \ea \right|
}{
\left| 
  \ba{cccc} 
    x&f_x(u_i)&f_x(u_j)&f_x(u_k)\\
    z&f_y(u_i)&f_y(u_j)&f_y(u_k)\\
    y&1&1&1\\
    w&f_w(u_i)&f_w(u_j)&f_w(u_k)
  \ea \right|
} .\een

Next, we list the necessary results by Murata et al. \cite{msy}
concerning birational maps on $\ponet$
whose space of initial conditions are given by 
blown-ups of $\ponet$ at generic 8 points on the smooth 
curve of degree $(2,2)$
\be \label{ponet} 
(x+y+1)(4xy-\frac{g_3}{\wp^3(2t)})&=&
\left(xy+x+y+\frac{g_2}{4\wp^2(2t)} \right)^2 .
\ee

Let $H_0$, $H_1$ and $E_i$ denote the total transform of
$x=c\in\pone$, that of $y=c'\in\pone$ and that of the exceptional divisor
generated by the blow up at the point $P_i$ respectively.
The Picard group and the canonical divisor are
\ben
\pic(X)&=& \mz H_0\oplus \mz H_1 \oplus \mz E_1\oplus \mz E_2\oplus \cdots 
\oplus \mz E_8\\
K_X&=&-2H_0-2H_1+E_1+E_2+E_3+E_4+E_5+E_6+E_7+E_8
\een
and the intersection numbers are given by
\ben &&H_i\cdot H_j=1-\delta_{i,j},\quad  
H_i\cdot E_j=0, \quad (E_i,E_j) = -\delta_{i,j} . \een
The root basis is given by
\ben &&\al_i=E_{7-i}-E_{8-i} \quad(i=0,1,\cdots,5),\\
&&\al_6=H_1-E_1-E_2,\quad \al_7=H_0-H_1,\quad \al_8=E_1-E_2 , \een

\begin{figure}[ht]
\hspace{2cm}
\begin{picture}(500,60)
\put(50,20){\line(1,0){20}}
\put(75,20){\line(1,0){20}}
\put(100,20){\line(1,0){20}}
\put(125,20){\line(1,0){20}}
\put(150,20){\line(1,0){20}}
\put(175,20){\line(1,0){20}}
\put(200,20){\line(1,0){20}}
\put(97.5,22.5){\line(0,1){20}}
\put(47.5,20){\circle{5}}
\put(72.5,20){\circle{5}}
\put(97.5,20){\circle{5}}
\put(122.5,20){\circle{5}}
\put(147.5,20){\circle{5}}
\put(172.5,20){\circle{5}}
\put(197.5,20){\circle{5}}
\put(222.5,20){\circle{5}}
\put(97.5,45){\circle{5}}
\put(47.5,30){\makebox(0,0){}}
\put(197.5,30){\makebox(0,0){}}
\put(47.5,10){\makebox(0,0){$\al_{7}$}}
\put(72.5,10){\makebox(0,0){$\al_{6}$}}
\put(97.5,10){\makebox(0,0){$\al_{5}$}}
\put(122.5,10){\makebox(0,0){$\al_{4}$}}
\put(147.5,10){\makebox(0,0){$\al_{3}$}}
\put(172.5,10){\makebox(0,0){$\al_{2}$}}
\put(197.5,10){\makebox(0,0){$\al_{1}$}}
\put(222.5,10){\makebox(0,0){$\al_{0}$}}
\put(115,45){\makebox(0,0){$\al_{8}$}}
\end{picture}
\caption[]{$E_8^{(1)}$ Dynkin diagram}
\end{figure}
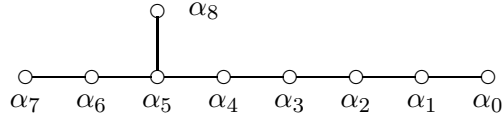

\noi and each action on the parameter space $({\bf u},t)$ becomes
\be && r_i: (u_{7-i},u_{8-i})\mapsto (u_{8-i},u_{7-i}), 
\qquad (i=0,1,\cdots,5) \nn\\
&& r_6:\left(\ba{cccc}u_1&u_2&u_3&u_4\\u_5&u_6&u_7&u_8\ea,t\right) \nn\\
&& \mapsto \left(
\ba{cccc}
u_1-3\lambda_{1,2}&u_2-3\lambda_{1,2}&u_3+\lambda_{1,2}&
u_4+\lambda_{1,2} \\
u_5+\lambda_{1,2}&u_6+\lambda_{1,2}&u_7+\lambda_{1,2}&
u_8+\lambda_{1,2}\ea,  t-\lambda_{1,2}
\right),  \nn \\
&& r_7: t \mapsto -t, \qquad r_8: (u_1,u_2)\mapsto (u_2,u_1), \label{acte8}
\ee
where $\lambda_{1,2}=\frac{1}{4}(2t+u_1+u_2)$.\\

\noi{\bf Proof of Theorem \ref{conserve}}\\
As mentioned above, it is sufficient to show for $r_7=r_{E-E_1-E_2-E_3-E_4}$
in case (i-1).
Let $G$ and $H$ denote the matrices corresponding to $(G)$ and $(H)$
of the pencil (\ref{penponet}). We write
$(\ol{x}:\ol{y}:\ol{z}:\ol{w}):=r_7(x:y:z:w)$.
Since the image of an element of the pencil 
$(\ol{x}:\ol{y}:\ol{z}:\ol{w})|_{{\bf x}^t (\al_0 G + \al_1 H){\bf x} =0}$
($(\al_0:\al_1)\in \pone$) is again an element of $|-\frac{1}{2}K_X|$
and therefore is again an element of the pencil (the intersection curve
does not move).
Hence, there exists $(\beta_0:\beta_1)\in \pone$ such that
$\ol{{\bf x}}^t (\beta_0 G + \beta_1 H)\ol{{\bf x}} =0$.
Since this correspondence defines an automorphism of the base space $\pone$
of the pencil, it is enough to show that 3 elements of the pencil
are preserved. Hence we may assume 
$\rank(\al_0G+\al_1H)=\rank(\beta_0G+\beta_1H)=4$.
In the following, we assume $(\al_0:\al_1) \neq (\beta_0:\beta_1)$ and lead a contradiction. 

\begin{lemma}\label{renor}
Assume $\rank(\al_0G+\al_1H)=4$. There exist 
$P\in PGL(4), v\in \mc^{\times}$ and $t'\in 
(\mc/(\mz+\mz \tau))\setminus \{0\}$ such that 
$$P^t(\al_0G+\al_1H)P=G, \quad P^tHP=v H(\wp(2t'))$$
and moreover there exist
$Q_1,Q_2 \in PGL(4), v_1,v_2\in \mc^{\times}$ and 
$c,d \in \mc\setminus \{0,1\}$ such that
\ben &Q_1^t(\al_0G+\al_1H)Q_1={\rm diag}(1,1,1,1), &Q_1^tHQ_1=v_1{\rm diag}
(0,1,c,d),\\
&Q_2^tH(\wp(2t'))Q_2={\rm diag}(1,1,1,1), 
&Q_2^tHQ_2=v_2{\rm diag}(0,1,c,d) .\een

\end{lemma}

\bp 
Since $\rank H=3$, similar to the proof of Theorem \ref{norp}, 
$G$ and $H$ can be transformed 
to $$G'={\rm Id.}, \quad H'={\rm diag}(0,a,b,1) \quad a,b,1
\mbox{ are different each other}$$
by $PGL(4)$. Here, $a$ and $b$ are functions of $g_2,g_3$ and $\wp(2t)$.
Conversely, if $a,b,1$ are different each other, there exist corresponding 
$g_2,g_3$ and $\wp(2t)$. On the other hand,
they are also transformed to 
$$(\al_0G+\al_1H)'={\rm Id.} \quad H'={\rm diag}(0,c,d,1)
 \quad c,d,1\mbox{ are different each other}$$
by $PGL(4)$.
Hence, there exist corresponding $g_2',g_3',\wp(2t')$ and 
$P\in PGL(4)$ such that
$$P: \al_0G+\al_1H, H \mapsto G, H(g_2',g_3',\wp(2t')) .$$
Here, $r_7$ defines a birational map between the intersection curves
and therefore the parameter $\tau=w_2/w_1$ does not change.
As Remark \ref{modu}, if we fix $w_1$ as $w_1=1$, only $t'$ can change,
which shows that $H'$ is a function depending only on $\wp(2t')$.\ep \\

For $(\al_0:\al_1)$ and $(\beta_0:\beta_1)$ we denote
$t'$ determined by Lemma \ref{renor} by $t_a$ and $t_b$ respectively.
First, we show $t_a \neq t_b$.  Assume $t_a=t_b$.
Since only $PGL(4)$ is needed for ${\rm diag}(0,1,c,d)$
when one normalizes the later form of the basis of pencil in the above lemma
to the form ${\rm diag}(0,1,1,1),{\rm diag}(1,1,\lambda,0)$,
there exist elements $J$ and $J'$ of pencil, where
$J\neq J', \rank J=\rank J'=3$, such that both pairs 
$J,H$ and $J',H$ can be transformed to the form 
${\rm diag}(0,1,1,1),{\rm diag}(1,1,\lambda,0)$
by $PGL(4)$.
Hence there exist elements $K={\rm diag}(1,1,1,0)$ and $K'$ of the pencil 
$\{s_0{\rm diag}(0,1,1,1)+s_1{\rm diag}(1,1,\lambda,0)\}$,
where $K\neq K', \rank K'=3$, such that
the pair $K,H$ can be transformed to the pair $K',H$ by $PGL(4)$.
The elements of this pencil in rank 3 are
$K={\rm diag}(0,1,1,1),{\rm diag}(1,1,\lambda,0),
K_1:={\rm diag}(1,0,\lambda-1,-1)$ and 
$K_2:={\rm diag}(-1,\lambda-1,0,\lambda)$.
Therefore, $K'$ should be $K_1$ or $K_2$.
Normalizing each by $PGL(4)$,
we have that $\lambda$ is $\lambda/(\lambda-1),(\lambda-1)/\lambda, 1-\lambda$
or $1/(1-\lambda)$.
If $\lambda \neq 2,\frac{1}{2}, 
\frac{1 \pm 3 I}{2}$, it does not coincides with $\lambda$.
Hence, if $\tau$ does not correspond to these $\lambda$,
we have $t_a \neq t_b$
(as shown by (\ref{xg2g3}), $\lambda$ does not depend on $\wp(2t)$)

Let $P_a,P_b$ denote the elements of $PGL(4)$ which give $t_a,t_b$
in Lemma \ref{renor}. We consider the map defined by
$P_b \circ r_7 \circ P_a^{-1}$.
Since the period map is conserved by $PGL(4)$,
we have
$\ol{u_i}=u_i-\lambda_1+c$ $(1\leq i \leq 4)$, $\ol{u_i}=u_i+\lambda_1+c$ 
$(5\leq i \leq 8)$, where $c$ is a constant.
On the other hand, the map
$$ {\bf x}^tG{\bf x}=0 \mapright{P_a^{-1}} {\bf x}^t(\al_0G+\al_1H){\bf x}=0
 \mapright{r_7} {\bf x}^t(\beta_0G+\beta_1H){\bf x}=0
\mapright{P_b} {\bf x}^tG{\bf x}=0$$
and Segr\'e map define a birational map on $\ponet$.

By blow up $\ponet$ at $\{P_i\}$ and $\{\ol{P_i}\}$ the map is lifted to
an isomorphism and included by the Weyl group of type $E_8^{(1)}$.
Since the Weyl group conserves 
$\chi_S(\delta)=-\sum_{i=1}^8 u_i$, we have $c=0$.
Hence $r_7$ can be written as
$$r_7=r_{H_0+H_1-E_1-E_2-E_3-E_4}=
r_6\circ r_7\circ r_5\circ r_8\circ r_4\circ r_5\circ r_6\circ 
r_5\circ r_4\circ r_8\circ r_5\circ r_7\circ r_6$$
by the root system of $E_8^{(1)}$, 
and we have $t_a=t_b$ form (\ref{acte8}), 
which contradicts $t_a\neq t_b$.

We have shown $(\al_0:\al_1)=(\beta_0:\beta_1)$
if $\tau$ does not correspond to $\lambda\neq 2,\frac{1}{2}$ or 
$\frac{1 \pm 3 I}{2}$.
When $\lambda= 2,\frac{1}{2}, \frac{1 \pm 3 I}{2}$
it can be shown by continuity of $\wp$ with respect to
$\tau$.\\
 \ep ~ {\bf Proof of Theorem \ref{conserve}}\\

\section{Conclusions and discussions}

In this paper, we defined the inner product for the Picard group
of varieties obtained by blow-ups at 8 points
in $\pth$ by means of the intersection numbers and the anti-canonical divisor
and showed that the symmetry group defined by means of the inner product
is the Weyl group of type $E_7^{(1)}$.
As in 2-dimensional case \cite{sakai}, if the configuration of points
is special, the symmetry may become smaller.

This method can be applied to other families of 3-dimensional
rational varieties. 

\begin{ex}
Let $X$ be a variety obtained by blow-ups at generic 6 points in
$\ponet \times \pone$ and let
$H_i$ and $E_i$ denote the total transform of divisor class of a plane 
such that one of its coordinate of $\ponet \times \pone$ is constant
and that of the exceptional divisor generated by a blow-up respectively.
The symmetry group $Gr(\{X\})$ becomes the Weyl group of type $E_7^{(1)}$
defined by the root system
\ben && \al_0=H_0-H_2, ~ \al_1=H_1-H_2,~ \al_3=H_2-E_1-E_2, \nonumber \\
     && \al_i=E_{i-1}-E_i ~(3\leq i \leq 6), ~~\al_7=E_1-E_2 ,\een
where the inner product is given by 
$(H_i,H_j)=1-\delta_{i,j},~ (H_i,E_j)=0, 
(E_i,E_j) = -\delta_{i,j}$.
This $X$ and the variety obtained of blow-ups at generic 8 points 
in $\pth$ are not isomorphic. Thus, the relation between these 2 Weyl groups
is not trivial.
\end{ex}

It may be worth commenting that the space of initial conditions
for Kajiwara-Noumi-Yamada's birational 
representation of the Weyl group of type $A_1^{(1)}\times A_2^{(1)}$   
\cite{kny} can be obtained when
the points of blow-ups are in special position
in the above example.
These examples are 3-dimensional but it is expected that
our method can be applied to 4(or more)-dimensional cases.\\

The other results are summarized as follows. \\

\noi $\cdot$ We normalized the configuration space by normalizing
the pencil of quadratic surfaces passing through 8 points 
($\in |-\frac{1}{2}K_X|$). The intersection curve is an elliptic 
curve in generic. 

\noi $\cdot$ In order to obtain concrete expression of 
the action of the Weyl group, we introduced a 3-dimensional
analogue of period map. 

\noi $\cdot$ We showed the action of the Weyl group preserves
each element of $|-\frac{1}{2}K_X|$ and therefore
it reduces to the action on $\ponet$.
The reduced action of the Weyl group of type $E_7^{(1)}$
is included by the action of the Weyl group of type $E_8^{(1)}$,
which is the symmetry of the family of generic Halphen surfaces. \\

{\noindent{\it Acknowledgment.}}
The author would like to thank K. Okamoto, H. Sakai, 
J. Satsuma, R. Willox, M. Eguchi, M. Murata and T. Tsuda 
for discussions and advice. 
The author also appreciates a Research Fellowship of the
Japan Society for the Promotion of Science for Young Scientists.

\end{document}